\documentclass[prl,twocolumn,showpacs,amsmath,amssymb,superscriptaddress]{revtex4-2}
\usepackage[english]{babel}
\usepackage{amsmath}
\usepackage{graphicx, nicefrac}
\usepackage{float}
\usepackage{siunitx}

\usepackage{booktabs}
\usepackage{breakurl}
\usepackage{color}
\usepackage [latin1]{inputenc}

\usepackage{dcolumn}
\usepackage{bm,graphicx}
\usepackage{etoolbox}

\usepackage{url}
\usepackage[colorlinks]{hyperref}
\usepackage[normalem]{ulem}
\usepackage{xcolor}
\usepackage[version=3]{mhchem} 
\usepackage{array}
\newcolumntype{x}[1]{>{\centering\arraybackslash\hspace{0pt}}p{#1}}
\usepackage{tabularray}
\usepackage{multirow}

\begin{document}


\title{Tuning THz magnons in a mixed van-der-Waals antiferromagnet}

\author{F.~Le Mardel\'e}
\email[]{florian.le-mardele@lncmi.cnrs.fr} 
\affiliation{LNCMI-EMFL, CNRS UPR3228, Univ. Grenoble Alpes, Univ. Toulouse, Univ. Toulouse 3, INSA-T, Grenoble and Toulouse, France}

\author{I.~Mohelsky}
\affiliation{LNCMI-EMFL, CNRS UPR3228, Univ. Grenoble Alpes, Univ. Toulouse, Univ. Toulouse 3, INSA-T, Grenoble and Toulouse, France}

\author{D.~Jana}
\affiliation{LNCMI-EMFL, CNRS UPR3228, Univ. Grenoble Alpes, Univ. Toulouse, Univ. Toulouse 3, INSA-T, Grenoble and Toulouse, France}

\author{A.~Pawbake}
\affiliation{LNCMI-EMFL, CNRS UPR3228, Univ. Grenoble Alpes, Univ. Toulouse, Univ. Toulouse 3, INSA-T, Grenoble and Toulouse, France}

\author{J.~Dzian}
\affiliation{LNCMI-EMFL, CNRS UPR3228, Univ. Grenoble Alpes, Univ. Toulouse, Univ. Toulouse 3, INSA-T, Grenoble and Toulouse, France}
\affiliation{Institute of Physics, Charles University, Ke Karlovu 5, Prague, 121 16 Czech Republic}

\author{W.-L.~Lee}
\affiliation{Institute of Physics, Academia Sinica, Nankang, Taipei 11529, Taiwan, Republic of China}

\author{K.~Raju}
\affiliation{Institute of Physics, Academia Sinica, Nankang, Taipei 11529, Taiwan, Republic of China}

\author{R.~Sankar}
\affiliation{Institute of Physics, Academia Sinica, Nankang, Taipei 11529, Taiwan, Republic of China}

\author{C.~Faugeras}
\affiliation{LNCMI-EMFL, CNRS UPR3228, Univ. Grenoble Alpes, Univ. Toulouse, Univ. Toulouse 3, INSA-T, Grenoble and Toulouse, France}

\author{M.~Potemski}
\affiliation{LNCMI-EMFL, CNRS UPR3228, Univ. Grenoble Alpes, Univ. Toulouse, Univ. Toulouse 3, INSA-T, Grenoble and Toulouse, France}
\affiliation{Institute of High Pressure Physics, PAS, Warsaw, Poland}
\affiliation{CENTERA, CEZAMAT, Warsaw University of Technology, Warsaw, Poland}

\author{M.~E.~Zhitomirsky}
\affiliation{Univ. Grenoble Alpes, CEA, IRIG, PHELIQS, 17 avenue des Martyrs, 38000 Grenoble, France}

\author{M.~Orlita}
\email[]{milan.orlita@lncmi.cnrs.fr} 
\affiliation{LNCMI-EMFL, CNRS UPR3228, Univ. Grenoble Alpes, Univ. Toulouse, Univ. Toulouse 3, INSA-T, Grenoble and Toulouse, France}
\affiliation{Institute of Physics, Charles University, Ke Karlovu 5, Prague, 121 16 Czech Republic}
\date{\today}

\begin{abstract}
Alloying stands out as a pivotal technological method employed across various compounds, be they metallic, magnetic, or semiconducting, serving to fine-tune their properties to meet specific requirements. Ternary semiconductors represent a prominent example of such alloys. They offer fine-tuning of electronic bands, the band gap in particular, thus granting the technology of semiconductor heterostructures devices, key elements in current electronics and optoelectronics. In the realm of magnetically ordered systems, akin to electronic bands in solids, spin waves exhibit characteristic dispersion relations, featuring sizeable magnon gaps in many antiferromagnets. The engineering of the magnon gap constitutes a relevant direction in current research on antiferromagnets, aiming to leverage their distinct properties for THz technologies, spintronics, or magnonics. In this study, we showcase the tunability of the magnon gap across the THz spectral range within an alloy comprising representative semiconducting van-der-Waals  antiferromagnets FePS$_3$ and NiPS$_3$. These constituents share identical in-plane crystal structures, magnetic unit cells and the direction of the magnetic anisotropy, but differ in the amplitude and sign of the latter. Altogether these attributes result in the wide tunability of the magnon gap in the Fe$_{1-x}$Ni$_x$PS$_3$ alloy in which the magnetic order is imposed by stronger, perpendicular anisotropy of iron.
\end{abstract}
\maketitle

%

The ongoing research on magnetic van-der-Waals materials is multifaceted, addressing fundamental issues related to novel topological and quantum phases of matter, while also exploring potential applications across various fields, thereby contributing the advancement of magnonics~\cite{BarmanJPCM21,AfanasievNM21,HortensiusNP21,LeendersNature24,YangNC24}. The pursuit of magnetic materials with continuously adjustable magnetic properties, especially the magnon spectrum, is a significant focus of current research efforts. This exploration holds promise for enabling the fabrication of multilayered structures that exhibit tailored multi-magnon gap excitations, precisely suited to meet specific demands.

The magnon energies (gaps) can be tuned by temperature or magnetic field, and to a lesser extent, by pressure, strain, and electric field. The charge accumulation is also an option when working with a-few-layer stacks or heterostructures. Another way of tuning, explored in this paper, is alloying of antiferromagnets~\cite{Rao:1992aa, Lee:2021aa, Selter:2021aa, Basnet:2021aa, Graham:2020aa, Masubuchi:2008aa,  Bhutani:2020aa, He:2003aa, Goossens:2000aa,  Shemerliuk:2021aa}. 
Even though the long-range magnetic order is -- strictly speaking -- excluded in such randomly organized alloys, the antiferromagnetism is often preserved and systematic trends in magnetic properties are found. In a few cases, tuning of the magnon gap has been reported, nevertheless, only in alloys with small mixing ratios~\cite{GeisPBC77,MischlerJPSSP81,LockwoodJPCSSP82,CiepielewskiPRB88}. 

Antiferromagnetic vdW crystals with tailored properties are often viewed as materials suitable for novel terahertz (THz) technologies. Future wireless communication with a high data throughput (6G technology and beyond) is one of them~\cite{HuangE23}. Exceptionally fast dynamics of the magnetic lattice~\cite{NemecNP18} and optically active magnon modes in the sub-THz and THz spectral ranges are their key features that may allow them to become active media in various THz devices~\cite{ToAPL24}, \emph{e.g.}, as fast optically-driven modulators of THz radiation~\cite{DeglInnocentiNP22,BelvinNC21}. The fine tunability is here required to match the magnon energy with the communication channels defined by the windows of atmosphere transparency. 

 \begin{figure*}[t]
	\includegraphics[width=1\linewidth]{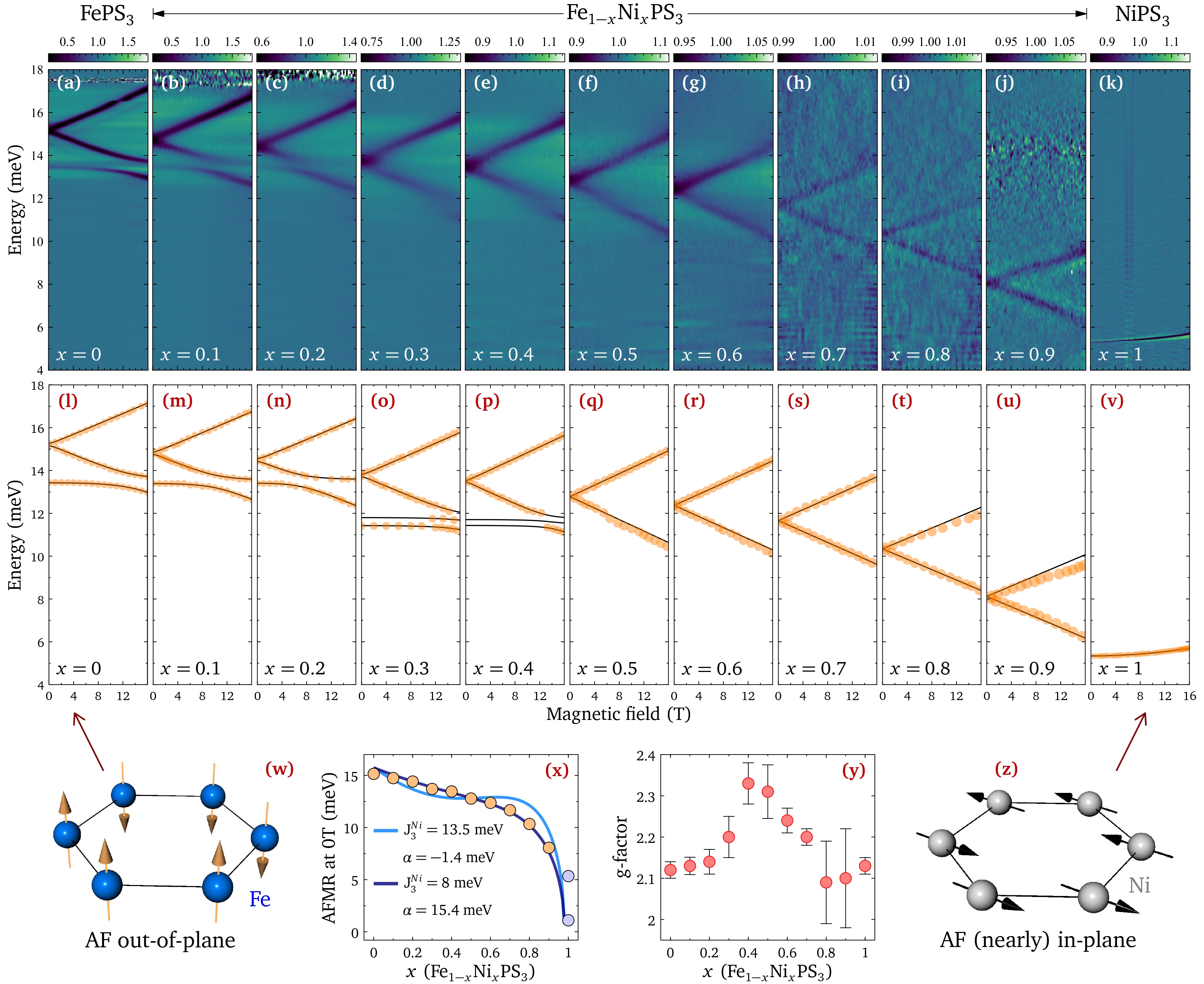}
\caption{Magneto-transmission data collected in the THz range on Fe$_{1-x}$Ni$_x$PS$_3$ alloys at $T = 4.2$~K for all decimal fractions of $x$ and plotted in the form of false-color maps (a-k). Each magneto-transmission spectrum is normalized by the averaged transmission over the explored range. The extracted positions of the AFMR branches were plotted in panels (l-v). Panels (w) and (z) illustrate the easy-axis and (nearly) easy-plane antiferromagnetic order in FePS$_3$ and NiPS$_3$, respectively. The extracted energies of the magnon modes at $B=0$ are shown in (x) together with theoretical modeling described in the main text. The energy of the lower magnon gap at 1.1~meV is taken from the recent magneto-Raman scattering study~\cite{JanaPRB23}. Deduced effective $g$ factors, average values obtained from both AFMR branches, are plotted in (y).}
	\label{fig1}
\end{figure*}

In this study, we conduct experimental investigations into alloys formed by mixing two layered antiferromagnets: FePS$_3$ and NiPS$_3$. These materials have identical crystal and magnetic lattices within the plane, featuring characteristic zigzag ferromagnetic chains aligned along the same crystallographic axis. 
In both cases, their magneto-crystalline anisotropy has an out-of-plane orientation, nevertheless, with a different strength and sign. Consequently, FePS$_3$ and NiPS$_3$ are identified as easy-axis and easy-plane antiferromagnets, respectively. Our findings demonstrate that by adjusting the molar fraction $x$ in the Fe$_{1-x}$Ni$_{x}$PS$_3$ alloy, the magnon energy can be continuously varied from 2 to 4 THz (8 to 16 meV). Remarkably, this dependence of the magnon energy can be interpreted in terms of varying the magnetic anisotropy, which maintains its perpendicular orientation up to relatively high nickel concentrations ($x\approx 0.9$).

The high-quality mixed crystals of Fe$_{1-x}$Ni$_x$PS$_3$~\cite{Lee:2021aa} were grown by a chemical vapor transport method using iodine as a transport agent. Initially, the polycrystalline powders were synthesized by the solid-state synthesis process under high vacuum conditions. The high-purity (5N) starting materials were weighted at a stoichiometric ratio and then sealed into the quartz tube with a diameter of 22~mm with 10$^{-3}$~Torr pressure. The mixed compounds were heated and grounded twice at 400 and 600$^\circ$C to make a single-phase compound. The 200~mg iodine was added into the polycrystalline samples and sealed by the tube dimension of 20$\times$22$\times$400 mm$^3$ with 10$^{-3}$~Torr pressure. The tube was kept for growth at a two-zone furnace with the temperature range of 600-700$^\circ$C for 200~h. After completing the growth process, the furnace temperature was reduced to room temperature at a rate of 2$^\circ$C/min. The quartz tube was broken inside an argon-filled glovebox and collected good-quality single crystals. For experiments on pure NiPS$_3$, a commercially available crystal was used.  

We carried out a series of low-temperature THz magneto-transmission experiments on Fe$_{1-x}$Ni$_{x}$PS$_3$ alloys with $B$ oriented along the $c$-axis (out-of-plane direction). The radiation from a mercury lamp was analyzed by a Bruker Vertex 80v Fourier-transform spectrometer, and delivered to the sample via light-pipe optics. The Fe$_{1-x}$Ni$_x$PS$_3$ samples -- with an effective irradiated area of several mm$^2$ and thickness of several hundred microns -- were kept at $T=4.2$~K in the helium exchange gas. The radiation was detected by a composite bolometer placed below the sample. The measured magneto-transmission spectra were normalized using the spectrum obtained by averaging over the whole range of $B$ explored which facilitates identifying $B$-dependent spectral features. The data collected on alloys with all decimal-fraction compositions, $x=0, 0.1\ldots 1$, are presented in Figs.~\ref{fig1}a-k, in the form of false-color plots. The extracted positions of $B$-dependent features are plotted in Figs.~\ref{fig1}l-v.   

To interpret the data, we first focus on the response of pure FePS$_3$ that is well understood thanks to several recent studies~\cite{McCrearyPRB20,LiuPRL21, VaclavkovaPRB21,ZhangCM21,WyzulaNL22}. These allow us to associate the observed feature with the $k=0$ magnon mode, corresponding to in-phase oscillations of parallel iron sublattices. Following the expectations for the classical antiferromagnetic resonance (AFMR) in easy-axis antiferromagnets~\cite{KittelPR51}, it symmetrically splits into two branches that evolve linearly with the applied magnetic field. Around $B=14$~T, the lower AFMR branch undergoes avoided-crossing behaviour due to coupling with an optical phonon. This is a signature that we observe a magnon-polaron rather than bare magnon modes~\cite{LiuPRL21,VaclavkovaPRB21,ZhangCM21}. 

With an increasing nickel content (Figs.~\ref{fig1}b-j), the overall AFMR-like character of the response is preserved. It is just the position of the magnon gap which redshifts monotonically, from 4~THz in pure FePS$_3$ down to 2~THz in Fe$_{0.1}$Ni$_{0.9}$PS$_3$. 
For concentrations up to $x=0.4$, the lower AFMR branch exhibits signs of magnon-phonon coupling. In alloys with a high nickel content ($0.9<x<1$), no magnon-like excitations were found in our data. In pure NiPS$_3$, the response is dramatically different and reflects the easy-plane antiferromagnetic order, reported in preceding studies~\cite{WildesPRB2015,WildesPRB22,JanaPRB23}. There, the degeneracy of the magnon mode is lifted even at $B=0$ and only the upper mode is seen at 5.3~meV in our data,  having approximately quadratic in $B$ dependence~\cite{JanaPRB23}. The lower magnon mode was observed in preceding  studies~\cite{MehlawatPRB22,AfanasievSA21,JanaPRB23}, but its energy of $1.1$~meV is too low to be resolved in our Fourier-transform experiments.

The observed behavior implies that, in a broad range of the nickel content, the Fe$_{1-x}$Ni$_x$PS$_3$ alloy behaves as an antiferromagnet with an out-of-plane easy axis. We assign this exceptional stability of the magnetic state to the fact that pure materials, FePS$_3$ and  NiPS$_3$, order into the same zig-zag spin structure with ferromagnetic arrays of parallel spins that alternate antiferromagnetically in the transverse direction. 
The only difference between the two is the orientation of ordered moments, which are orthogonal to the $ab$ plane in FePS$_3$, see Fig.~\ref{fig1}w, but nearly in-plane for NiPS$_3$, see Fig.~\ref{fig1}z. A  sensitivity of the moment direction in an alloy to even a tiny concentration of iron can be assigned to its particularly large single-ion anisotropy. It is worth mentioning that in other similar alloys, such as Fe$_{1-x}$Mn$_{x}$PS$_3$, where the pure compounds have different magnetic arrangement (zig-zag versus N\'eel), a more complex evolution of magnetic properties has been reported~\cite{Masubuchi:2008aa}. 

To describe the long-range magnetic order established in Fe$_{1-x}$Ni$_x$PS$_3$ quantitatively, we introduce a global antiferromagnetic order parameter ${\bf l}$, $|{\bf l}|=1$. Further, we consider the single-ion magnetic anisotropy in the form of $-DS_z^2$ for both materials, neglecting an order-of-magnitude weaker anisotropy within the plane reported in NiPS$_3$~\cite{WildesPRB22}. The magnetic anisotropy energy normalized per mole of transition metal ions then reads:
\begin{equation}
E_{\rm an} = -N_A\cos^2\theta\,\bigl[D_{\rm Fe} S_{\rm Fe}^2(1-x) + D_{\rm Ni} S_{\rm Ni}^2\,x\bigr],
\label{Ean}
\end{equation}
where $\theta$ is the angle between ${\bf l}$ and the $c$-axis, $N_A$ stands for the Avogadro constant. $D_{\rm Fe} = 2.66$~meV~\cite{LanconPRB16} and $D_{\rm Ni} =  -0.21$~meV~\cite{WildesPRB22} are single-ion anisotropy constants for two magnetic ions with the respective spins $S_{\rm Fe} = 2$ and $S_{\rm Ni} = 1$. We find that the out-of-plane orientation of ordered moments is energetically favourable  (i.e., $E_{\rm an} <0$) up to a relatively high nickel concentration: $x_c = 4/[4-D_{\mathrm{Ni}}/D_{\mathrm{Fe}}] \approx 0.98$, in agreement with our experimental findings.

The magnon gaps, accessed directly in our magneto-optical experiments, can be computed using the hydrodynamic spin-wave theory~\cite{HalperinPR69,AndreevSPU80}. The consideration is based on the Lagrangian for a collinear antiferromagnet, described by the order parameter $ {\bf l}$:
\begin{equation}
{\cal L} = \frac{\chi_\perp}{2} (\partial_t {\bf l})^2 - E_{\rm an}, 
\label{Lag}
\end{equation}
where $\chi_\perp$ is the transverse susceptibility. The 
magnetic anisotropy energy (\ref{Ean}) is here expressed as $E_{\rm an}= -\frac{a}{2}\, l_z^2$, where $l_z=\cos\theta$ and $a= 2N_A[D_{\rm Fe} S_{\rm Fe}^2(1-x) + D_{\rm Ni} S_{\rm Ni}^2\,x]$.

In the easy-axis case ($a>0$), the antiferromagnetic vector is oriented along the $c$-axis. Then, the equation of motion for (\ref{Lag}) yields two degenerate magnon modes with the gap:
\begin{equation}
\Delta_{1,2} = \sqrt{a/\chi_\perp}\,.
\label{gap}
\end{equation}
For the easy-plane anisotropy ($a<0$), the order parameter ${\bf l}$ lies in the basal plane and two magnon gaps read $\Delta_1=0$, $\Delta_2 = \sqrt{|a|/\chi_\perp}$. It is worth noting that the calculated energies of magnon gaps are in line with $k=0$ gaps obtained using the standard linear spin-wave theory applied to pure materials,  FePS$_3$~\cite{WyzulaNL22,LeMardelePRB24} and NiPS$_3$~\cite{WildesPRB22}.

To compute magnon gaps in the magnetically ordered alloy, we use the coherent potential approximation (CPA) and find the parameters $a$ and $\chi_\perp$ that enter Eq.~\ref{gap}.
The CPA approximation is well established for electrons in solids~\cite{ElliotRMP74,ChenPRL78,MouradEPJB12}, including their magnetic properties~\cite{Harrispss72,HolcombJPC74,HalleyJPC78}.  
In the easy-axis case ($a>0$), we obtain the following expression
for the twice degenerate magnon gap (\ref{gap}):
\begin{equation}\label{gapA}
\begin{split}
\Delta_{1,2}^2 = 4\overline{DS^2} \bigl[3\overline{J_3S^2} + \overline{J_1S^2} + 4\overline{J_2S^2} + \overline{DS^2}\bigr]/\overline{S},
\end{split}
\end{equation}
where the averaged microscopic parameters depend on the composition $x$ of the alloy:
\begin{gather} \label{average}
\overline{S} =  S_{\rm Fe}(1-x) + S_{\rm Ni}\,x\\ 
\label{anisotropy}
\overline{DS^2} =  D_{\rm Fe}S_{\rm Fe}^2(1-x) + D_{\rm Ni} S_{\rm Ni}^2\,x\\
\begin{split}\label{exchange}
    \overline{J_nS^2} = J_n^{\rm FeFe} S_{\rm Fe}^2(1-x)^2 + J_n^{\rm NiNi} S_{\rm Ni}^2\, x^2 +\\ +2J_n^{\rm FeNi} S_{\rm Fe}S_{\rm Ni}(1-x)x.     
\end{split}
\end{gather}

For FePS$_3$ and NiPS$_3$, there exists a solid set of microscopic parameters deduced from neutron scattering experiments, see Tab.~~\ref{Tab}. These allow us to estimate the averaged values of microscopic parameters $\overline{S}$ and $\overline{DS^2}$ needed to calculate the magnon gap for a given composition $x$. The situation is, however, more complex for the average exchange constants $\overline{J_nS^2}$. These also depend on a priori unknown strength of the exchange coupling between pairs of iron and nickel $J_n^{\mathrm{FeNi}}$. Fortunately, the form of Egs.~\ref{gapA} and \ref{exchange} allows us to introduce a single effective Ni-Fe exchange constant $\alpha= 2S_{\rm Fe}S_{\rm Ni}[J_1^{\rm FeNi}+4J_2^{\rm FeNi}+3J_3^{\rm FeNi}]$ that we use as a fitting parameter.

\begin{table}[t]
    \centering
    \renewcommand{\arraystretch}{1.6}
    \begin{tabular}{|x{15mm}|x{20mm}|x{20mm}|}
 \hline
        & {\textbf{FePS$_3$}}~\cite{LanconPRB16} & {\textbf{NiPS$_3$}}~\cite{WildesPRB22}\\   
\hline
\hline
         $J_1$~(meV)& -2.92 & -2.6 \\
         $J_2$~(meV)&  0.08  & 0.2\\
         $J_3$~(meV)& 1.92 &  13.5 \\
         $D$~(meV)& 2.66 & -0.21  \\
\hline  
\end{tabular}
\caption{Microscopic parameters of FePS$_3$ and NiPS$_3$ deduced from fits of neutron scattering experiments in Refs.~\cite{LanconPRB16,WildesPRB22}.}
    \label{Tab}
\end{table}

The magnon gap calculated using the above introduced hydrodynamical model is compared with experimentally extracted values in Fig.~\ref{fig1}x. Several observations can be made: (i) reasonable semi-quantitative agreement can be obtained (light blue line) when $\alpha$ is kept as the only free parameter, with the best agreement for $\alpha = -1.4$~meV; (ii) a closer analysis shows that quantitative agreement, with a truly monotonic $\Delta_{1,2}(x)$ dependence, can be achieved when the $J_3^{\mathrm{Ni}}$ exchange constant, particularly strong for nickel, is ad hoc reduced (dark blue curve), thus serving as an additional fitting parameter; (iii) the observed redshift of the magnon mode with $x$ is, to a great extent, due to effective tuning of the out-of-plane (perpendicular) magnetic anisotropy, cf. Eqs.~\ref{gapA} and \ref{anisotropy}, concluded in preceding magnetization studies~\cite{Rao:1992aa,Lee:2021aa} and (iv)
no AFMR signal was observed in alloys with nearly or completely vanishing magnetic anisotropy ($0.9<x<1$). An interesting question arises for the critical concentration $x_c \approx 0.98$ for which $E_{\mathrm{an}}$ in Eq.~\ref{Ean} vanishes. 
For this value of $x$, an approximate Heisenberg symmetry is restored in the hydrodynamic CPA approximation. Such a symmetry, of course, appears only macroscopically. At the microscopic scale, a highly nontrivial spin texture should be present, which may lead to a spin-glass type state at $T=0$. In any case, the ordering temperature must be strongly suppressed for samples with the concentration around  $x_c$.

It is instructive to put the last two points in a broader context of recent research on magnetic materials. Over the past few years, there was a considerable interest in the perpendicular magnetic anisotropy, primarily motivated by possible applications in fast read-write and laterally-dense data storage~\cite{DienyRMP17}. This concerns a wide class of magnetic materials, such as thin layers of bulk ferromagnets, interfaces of magnetic and non-magnetic systems, but also antiferromagnets, including synthetic ones, see, \emph{e.g.}, Refs.~\cite{NakajimaPRL98,LiuPRL15,ZhuravlevAPL18,HanPRA23}. In our case, our data show a possibility to effectively tune, on demand, the strength of the perpendicular anisotropy just by mixing two sibling materials with the same magnetic lattice, with the same orientation, but different strength and sign of the magnetic anisotropy.   

The dependence of the effective $g$ factor, deduced from the slope of the AFMR branches, is another interesting result of our experiments (see Fig.~\ref{fig1}y). The $g$ factor for pure compounds only slightly exceeds the value expected for a bare electron ($g_{\mathrm{Fe}}\approx g_{\mathrm{Ni}}\approx 2.1$), but the $g$ factor in the alloys visibly increases, reaching its maximum around $g\approx 2.35$ at $x\approx 0.5$. While any departure of the $g$ factor from the free electron value must be associated with a spin-orbit coupling, we do not see any clear mechanism responsible for the observed enhancement, notably in a compound with relatively light atoms, and therefore, relatively weak spin-orbit interaction. We speculate about inhomogeneity in the alloy that may create local distortions of sulfur-octahedra around magnetic ions. As a result, the crystal field levels and effective $g$ factors may got modified in comparison to pure compounds.

To conclude, we have experimentally studied the antiferromagnetic resonance in the mixed vdW crystal Fe$_{1-x}$Ni$_{x}$PS$_3$. Our data show that the long-range magnetic order exists even in mixed crystals with a random distribution of magnetic atoms lacking translational symmetry. The observed redshift of the AFMR mode, across the THz range, with the increasing nickel concentration demonstrates a possibility to tune widely, and on demand, the magnon energy just by choosing an appropriate mixing ratio of vdW antiferromagnets. Alloying antiferromagnetic materials thus emerge as a technologically pertinent approach for precisely tailoring their properties. It may facilitate the fabrication of magnonic structures with a reduced dimensionality, similar to the confinement of electrons in quantum wires or dots. 

\vspace{3mm}
{\large\noindent\textbf{Acknowledgement}}
\vspace{1mm}

M.E.Z.\ acknowledges support by the ANR,  France,  Grant No.~ANR-19-CE30-0040. The work was supported by the Czech Science Foundation, project No. 22-21974S. The work has been supported by the exchange programme PHC ORCHID (50852UC). R.S. acknowledges the financial support provided by the Ministry of Science and Technology in Taiwan under project numbers NSTC-111-2124-M-001-009; NSTC-110-2112-M-001-065-MY3; AS-iMATE-113-12. M.P. acknowledges support from the European Union (ERC TERAPLASM No. 101053716) and the CENTERA2, FENG.02.01-IP.05-T004/23 project funded within the IRA program of the FNP Poland, co-financed by the EU FENG Programme.


%

\end{document}